\documentclass[aps,prl,reprint,floatfix,superscriptaddress]{revtex4-1}
\usepackage{amsmath,amssymb}
\usepackage{graphicx}
\usepackage[utf8]{inputenc}
\usepackage{hyperref}
\usepackage{color}
\hypersetup{colorlinks=true,citecolor={blue},linkcolor={blue},urlcolor={blue}}
\usepackage{natbib}
\newcommand{\unit}[1]{\ensuremath{\, \mathrm{#1}}}
\newcommand{\ud}{\mathrm{d}}
\begin{document}

%\preprint{}

\title{The Non-linear Optical Spin Hall Effect and Long-Range Spin Transport in Polariton Lasers}% Force line breaks with \\

\author{E. Kammann}

\affiliation{%
 School of Physics and Astronomy, University of Southampton, Southampton, SO17
 1BJ, United Kingdom
}%
\author{T.C.H. Liew}
\affiliation{
School of Physical and Mathematical Sciences, Nanyang Technological University, 637371, Singapore}

\author{H. Ohadi}%
\author{P. Cilibrizzi}%

\affiliation{%
 School of Physics and Astronomy, University of Southampton, Southampton, SO17
 1BJ, United Kingdom
}%

\author{P. Tsotsis}%
\affiliation{Microelectronics Research Group, IESL-FORTH, P.O. Box 1527, 71110
Heraklion, Crete, Greece}

\author{Z. Hatzopoulos}
\affiliation{Microelectronics Research Group, IESL-FORTH, P.O. Box 1527, 71110
Heraklion, Crete, Greece}
\affiliation{Department of Physics, University of Crete, 71003 Heraklion, Crete, Greece}

\author{P.G. Savvidis}%
\affiliation{Microelectronics Research Group, IESL-FORTH, P.O. Box 1527, 71110
Heraklion, Crete, Greece}
\affiliation{Department of Materials Science and Technology, University of Crete, P.O.
Box 2208, 71003 Heraklion, Greece}

\author{A.V. Kavokin}%
\affiliation{%
 School of Physics and Astronomy, University of Southampton, Southampton, SO17
 1BJ, United Kingdom
}%
\affiliation{%
Spin Optics Laboratory, St-Petersburg State University, 1, Ulianovskaya,
St-Peterbsurg, 198504, Russia
}%

\author{P.G. Lagoudakis}
\email[correspondence address: ]{pavlos.lagoudakis@soton.ac.uk}
\affiliation{%
 School of Physics and Astronomy, University of Southampton, Southampton, SO17
 1BJ, United Kingdom
}%

\date{\today}

\begin{abstract} We report on the experimental observation of the non-linear
analogue of the optical spin Hall effect under highly non-resonant circularly
polarized excitation of an exciton polariton condensate in a GaAs/AlGaAs
microcavity. The circularly polarized polariton condensates propagate over
macroscopic distances while the collective condensate spins coherently precess
around an effective magnetic field in the sample plane performing up to four
complete revolutions.  \end{abstract}


\maketitle

Semiconductor microcavities in the strong coupling regime are excellent
candidates for designing novel ``spinoptronic" devices due to their strong
optical non-linearity, unusual polarization properties and fast spin-dynamics.
The first steps towards the fabrication of spin-based switching have  been
recently demonstrated
\cite{lagoudakis_stimulated_2002,amo_exciton|[ndash]|polariton_2010,adrados_motion_2011}.
An important goal for the development of integrated devices is coherent spin
transport. Being neutrally charged, exciton-polaritons have a significantly
smaller scattering cross section with atomic cores, than electrons in a metal.
Frictionless flow, which is one of the characteristics of a superfluid, has been
recently demonstrated in polaritons \cite{amo_superfluidity_2009}. Fabrication
of high finesse microcavities has allowed ballistic polariton propagation and
long-range order extending over macroscopic distances far beyond the excitation
area \cite{wertz_spontaneous_2010}. Here we show the coherent transport of the
spin vector in propagating polariton condensates. We observe ballistic
propagation of spin polarized polaritons over distances of a few hundred
microns.  The observed non-dissipative long-range spin transport is caused by
mass transport of exciton-polaritons which distinguishes the phenomenon we
observe from spin superfluidity reported for  $^3He$
\cite{borovik-romanov_long-lived_1984}, where the spin transport is decoupled
from the mass transport.

\begin{figure}  
\includegraphics[width=0.5\textwidth]{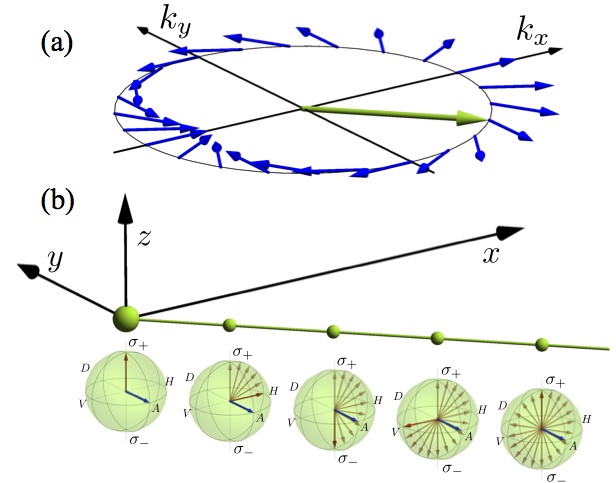}%
\caption{Polaritons experience an effective magnetic field that depends on their
in-plane $k$-vector. The effective magnetic field (blue arrows) is plotted at each
point on a ring in the momentum space (a). (b) A polariton injected at the
center of the excitation spot with an initial spin $\sigma_z = +1$ (marked by red
arrows) and moving outwards (along the green vector) experiences a spin
precession along the effective magnetic field (shown by the blue arrow in (a)).
The precession of the pseudospin vector in the Poincar\'{e} sphere at points where
$\vec{x}\cdot\vec{k}=\pi/2$, $\pi$, $3\pi/2$, $2\pi$ is shown.}  \label{fig:spheres}
\end{figure}

The polarization state of exciton-polaritons can be described within the
pseudospin formalism \cite{kavokin_quantum_2004}. Polaritons possess a spin with
two possible projections on the structural growth axis of the microcavity. The
polarization of the emitted light gives direct access to the pseudospin state,
which is fully characterized by the 4 component Stokes vector $\vec
s=(s_0,s_x,s_y,s_z)$. Here $s_0$ is the total degree of polarization and
$s_{x,y,z}=(I_{H,D,\circlearrowright}-I_{{V,A,\circlearrowleft}})/I_{tot}$.
$I_{H,D,\circlearrowright}$ and $I_{{V,A,\circlearrowleft}}$ are the measured
intensities in the horizontal and vertical, diagonal and anti diagonal and the
two circular polarization components and $I_{tot}$ is the total emission
intensity. The transverse electric and magnetic mode splitting in the
microcavity acts as a directionally dependent effective magnetic field (FIG.~\ref{fig:spheres} (a)), which causes the precession of the pseudospin for polaritons (FIG.~\ref{fig:spheres} (b)). This phenomenon being the optical analogue of the spin Hall \cite{dyakonov_current-induced_1971} effect was first predicted by Kavokin and coworkers for resonant Rayleigh scattering \cite{kavokin_optical_2005} and was experimentally observed in a semiconductor microcavity in the strong coupling regime \cite{leyder_observation_2007} and a purely photonic cavity \cite{maragkou_optical_2011} under resonant injection of polaritons and photons. In this Letter we demonstrate the non linear optical spin Hall effect occurring by the formation of a polariton ring condensate in momentum space from incoherently injected polaritons.

We use a $5\lambda/4$ AlGaAs/GaAs microcavity with a Rabi splitting of
$\sim9\unit{meV}$ and a cavity photon lifetime of $\sim9\unit{ps}$
\cite{tosi_sculpting_2012}. A circularly polarized continuous wave excitation
was tuned to a reflection minimum of the Bragg mirror outside the high reflectivity
region and focused to a $\sim 5 \unit{\mu m}$ diameter spot through a $0.2
\unit{NA}$ objective. All experiments were performed at
$\sim 7\unit{K}$ using a cold finger cryostat. The excitation laser was
intensity modulated using an acousto optic modulator at 10 kHz with a $5\%$ duty
cycle to reduce sample heating. The emission was spectrally separated from the
excitation laser and imaged onto a water-cooled CCD or sent to a $300 \unit{mm}$ imaging spectrometer. Calibrated waveplates and a polarizer were positioned in the detection path to analyse the polarization.

\begin{figure}
\includegraphics[width=0.5\textwidth]{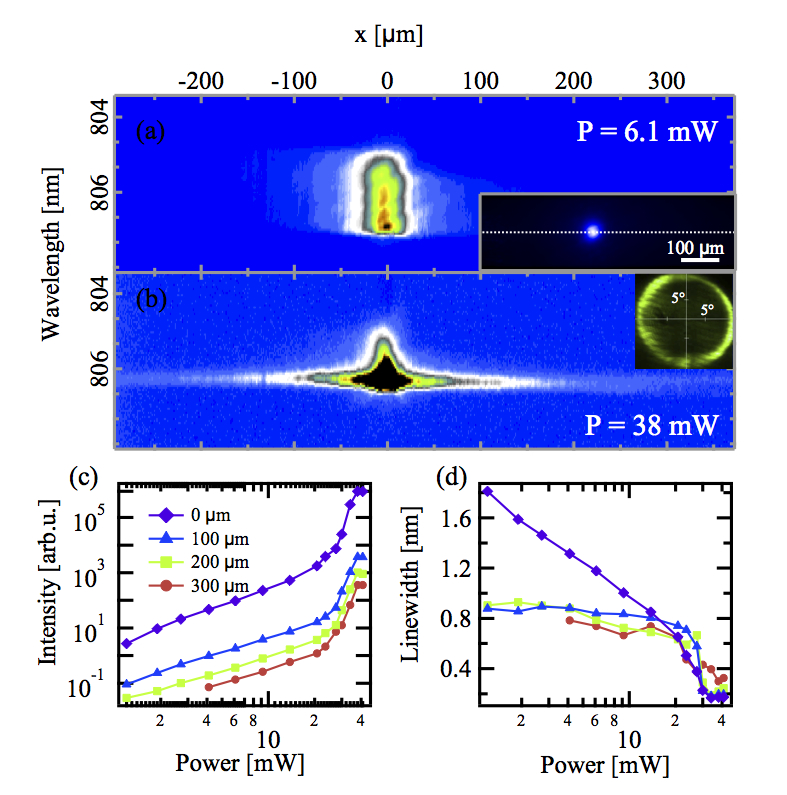}%
\caption{Energy resolved emission of a cross section through the image of the spot (inset in a) below (a) and above threshold (b). The inset in (b) shows the far field emission above threshold. (c) Emission intensity at various distances from the excitation spot show a superlinear increase at a threshold of about $25 \unit{mW}$. (d) The linewidth as a function of power rapidly drops at the threshold (color legend in (c)).} \label{fig:pd}
\end{figure}

When the sample is excited with a sufficiently small excitation spot, ($\lesssim
10\unit{\mu m}$), the repulsive interaction causes a strong blueshift of the condensate and radial ballistic propagation
of condensed polaritons out of the excitation spot. FIG.~\ref{fig:pd} shows the
energy-resolved emission of a cross section through the excitation spot for
below (a) and above (b) threshold power. The far-field emission forms a ring
above the threshold in reciprocal space (inset in FIG.~\ref{fig:pd} (b)). The collected emission spectra at different distances from the excitation area show a superlinear increase (c) and a rapidly dropping linewidth (d) at the photoluminescence threshold as far as $300\unit{\mu m}$ away from the excitation spot. These features are characteristic for lasing, condensation and buildup of coherence.

\begin{figure*}
\includegraphics[width=1\textwidth]{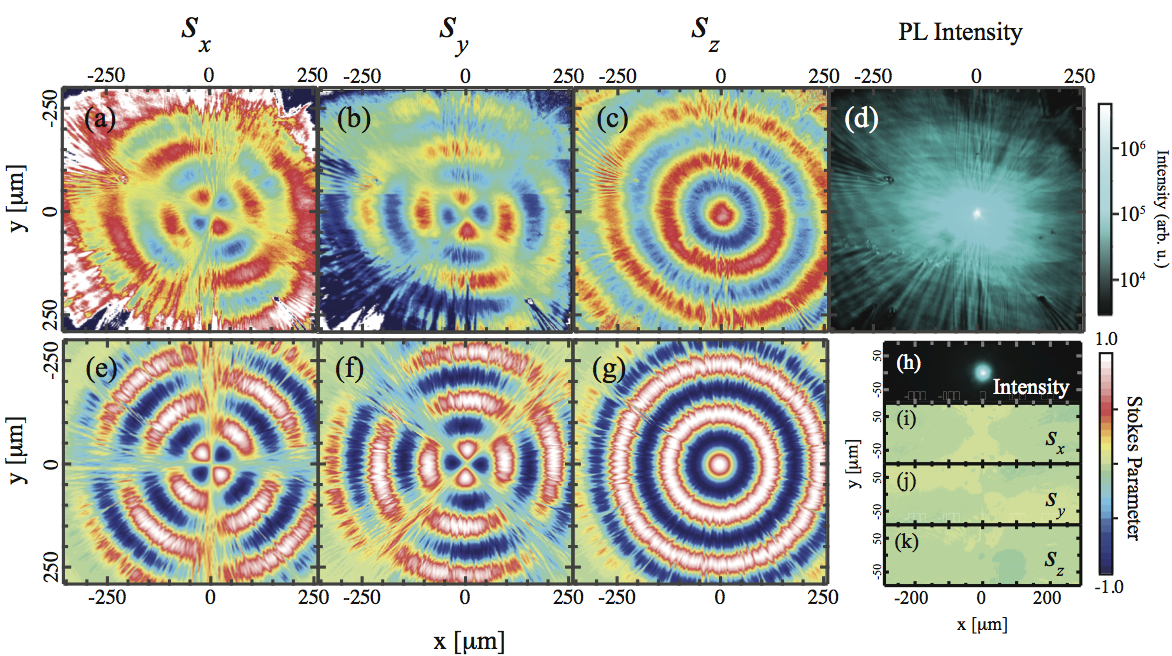}%
\caption{(a) Experimental data (a-c) and numerical simulations (e-g) of the
Stokes parameters of the emission of a polariton condensate non-resonantly
excited with circularly polarized laser at $2 \times P_{thr}$, $s_x$ (a,e) $s_y$
(b,f) $s_z$ (c,g). (d) Emission intensity on a logarithmic color scale. Below
the $P_{thr}$ the effect cannot be observed, due to the broad $k$-space distribution (h-k).} \label{fig:rings}
\end{figure*}

In reference \cite{ohadi_spontaneous_????} we showed that the spin of the
excitation laser is conserved to a certain degree under non resonant circularly
polarized excitation, whilst the phase correlation between the spin up and spin
down polaritons is lost. We utilize this phenomenon to non-resonantly form
polariton condensates with a collective spin state and study their long-range
spin transport. The transverse electric and the transverse magnetic photon mode
splitting causes the rotation of the spin as the polaritons propagate through
the sample, due to the optical spin Hall effect \cite{kavokin_optical_2005}.
FIG.~\ref{fig:rings} (a) shows the integrated emission mapped in the near field.
The linear components of the Stokes vector ($s_x$ and $s_y$) exhibit a cartwheel
pattern (FIG.~\ref{fig:rings} (b) and FIG.~\ref{fig:rings} (c)). The circular
component (FIG.~\ref{fig:rings} (d)) reveals up to four revolutions of the pseudospin around the effective magnetic field within the polariton lifetime and a circular symmetric ring pattern. 

To demonstrate the non-linearity of the effect the same measurement was
performed below the threshold. Here polaritons are evenly distributed in
$k$-space and therefore experience different effective magnetic fields. Fig.~\ref{fig:rings} (h) depicts the intensity below threshold. The Stokes components $s_x$, $s_y$ and $s_z$ are displayed in FIG.~\ref{fig:rings} (i),(j) and (k) respectively. In case of a linearly polarized excitation laser the ring condensate is randomly polarized within the excitation area \cite{baumberg_spontaneous_2008,ohadi_spontaneous_????} and no spin pattern was observed. 
Theoretically, the spatial dynamics of polariton condensates is described by a Gross-Pitaevskii type equation for the polariton field \cite{carusotto_probing_2004}, which should be coupled to a reservoir of hot excitons that are excited by the non-resonant pump \cite{wouters_excitations_2007}. The Gross-Pitaevskii equation is generalized to include the polarization degree of freedom \cite{shelykh_polarization_2006} of polariton condensates:
\begin{align}
&i\hbar \frac{\ud\Psi_{\sigma}(x,t)}{\ud t}=\bigg(-\frac{\hbar^2}{2m}\nabla^2-i\hbar\frac{\gamma}{2}+\alpha\vert\Psi_{\sigma}(x,t)\vert^2\nonumber\\
&+\big(g_R+i\hbar\frac{r}{2}\big) n_\sigma(x,t)+\hbar GP_\sigma(x,t)+V(x)\bigg)\Psi_{\sigma}(x,t)\nonumber\\
&+\frac{\Delta_{LT}}{k^2_{LT}} \bigg(i\frac{\partial}{\partial x}+\sigma\frac{\partial}{\partial y}\bigg)^2 \Psi_{-\sigma}(x,t)
\end{align}
$\Psi_{\sigma}(x,t)$ represents the mean field of polaritons, with $\sigma=\pm$ representing the spin of polaritons. We approximate the polariton dispersion as parabolic, with effective mass $m$, which is valid since the observed energies and in-plane wavevectors in the experiment lie in the parabolic part of the dispersion. $\gamma$ represents the polariton decay rate. $\alpha$ is the polariton-polariton interaction strength, which was assumed spin independent for simplicity (the interactions between polaritons with opposite spins are much weaker \cite{ciuti_role_1998}. $n_\sigma (x,t)$ is the density of the hot exciton reservoir, which may be polarized depending on the pump polarization but is assumed incoherent. $g_R$ represents the effect of repulsive interactions between the reservoir and polaritons (also assumed negligible for oppositely polarized spins) and $r$ is the condensation rate, representing the process where hot excitons condense into polaritons. An additional pump-induced shift is described by the interaction constant $G$ where $P_\sigma (x,t)$ is the spatial pump distribution \cite{wouters_excitations_2007}. $V(x)$ represents the static disorder potential typical in semiconductor microcavities, which is chosen as a random Gaussian correlated potential \cite{leyder_observation_2007}. The last term represents the presence of longitudinal-transverse splitting of the polariton modes \cite{panzarini_exciton-light_1999,kavokin_optical_2005}, which is assumed to increase with the square of the in-plane wavevector (parabolic approximation). $\Delta_{LT}$ defines the longitudinal-transverse splitting at $k=k_{LT}$.\\The evolution of the hot exciton density is given by the rate equation:
\begin{align}
\frac{\ud n_\sigma(x,t)}{dt}=\big(-\Gamma+\Psi + r\vert\Psi_\sigma(x,t)\vert^2\big)n_\sigma(x,t)+P_\sigma(x,t),
\end{align}
where $\Gamma$ is the reservoir decay rate. We consider a circularly polarized continuous wave pump. Starting from a random initial condition the time evolution of the system can be calculated numerically until a steady state is reached, which is independent of the initial condition. FIG.~\ref{fig:rings}(e)-(f) show the distribution of the calculated Stokes vectors in space \footnote{Parameters: $m=7\times 10^{-5}$ of the free electron mass, $\Delta_{LT}=0.05 \unit{meV}$, $k_{LT}=2.05\unit{\mu m^{-1}}$,$\alpha=2.4\mu \unit{eV}\:\unit{\mu m^2}$,$\gamma=0.2\unit{ps^1}$, $\Gamma=10\gamma$, $\hbar r=0.1\unit{meV}\:\unit{\mu m^{-2}}$, $G=0.03\unit{\mu m^2}$.The disorder potential was generated with 0.05 meV root mean squared amplitude and $1.5\unit{\mu m}$ correlation length. The pump intensity was chosen to match the experimentally measured blueshift of the polariton condensate.}.

Although we have included a disorder potential in our theoretical model, the
Rayleigh scattering of polaritons with disorder is not necessary for the
observation of the multiple rings and cartwheel structure of the polarization in
space. This is in contrast to the original demonstration of the optical spin
Hall effect
\cite{kavokin_optical_2005,leyder_observation_2007,maragkou_optical_2011} where
Rayleigh scattering was required to populate a ring in reciprocal space. In our
case, polaritons condense at the laser spot position with a blue-shifted energy
due to their interactions with uncondensed hot excitons. While these hot
excitons experience a limited diffusion, polaritons ballistically fly away from
the laser spot, converting this interaction energy into kinetic energy. The
kinetic energy is characterized by the non-zero wavevector of polaritons, which
due to the circular symmetry of the excitation corresponds to a ring in
reciprocal space.

The disorder potential does however have a noticeable effect on the fine structure of the polarization in space. Without disorder, similar calculations reveal smooth rings with perfect circular symmetry in the $s_z$ distribution (and smooth profiles with order 2 rotational symmetry in the $s_x$ and $s_y$ distributions). The addition of disorder gives the rings a noticeable texture and breaks the perfect circular symmetry (leaving only approximately symmetrical distributions). The observed texture is similar to that recorded experimentally. Although this indicates that a small amount of scattering with disorder is present, it is clear that we are in a regime of weak scattering or it would not be possible to observe such clear polarization patterns over such distances.

In conclusion we have experimentally demonstrated the non-linear optical spin Hall effect in a polariton condensate, with remarkable agreement with the theoretical prediction. The non-resonant circular excitation allows the excitation of a ring in reciprocal space without the need for large amounts of disorder. Polariton spins propagate ballistically over a $300\unit{\mu m}$ distance with minimal scattering and minimal loss of spin information. This record confirms the great potential of semiconductor microcavities for the fabrication of spinoptronic devices.  %The results of the spatial distributions of the Stokes parameters are given in FIG. \ref{fig:theory}\begin{figure}
%\begin{acknowledgments}
%\end{acknowledgments}

\bibliography{spinhall}

\end{document}